\documentclass[prb,twocolumn,showpacs,preprintnumbers,amsmath,amssymb]{revtex4}

\usepackage{graphicx}
\usepackage{dcolumn}
\usepackage{bm}

\begin{document}

\newcommand*{\cm}{cm$^{-1}$\,}


\title{Optical study on the spin-density wave properties
in single crystalline Na$_{1-\delta}$FeAs}

\author{W. Z. Hu}
\author{G. Li}
\author{P. Zheng}
\author{G. F. Chen}
\author{J. L. Luo}
\author{N. L. Wang}

\affiliation{Beijing National Laboratory for Condensed Matter
Physics, Institute of Physics, Chinese Academy of Sciences,
Beijing 100190, China}

\begin{abstract}
We report an optical investigation on the in-plane charge dynamics
for Na$_{1-\delta}$FeAs single crystal. A clear optical evidence
for the spin-density wave (SDW) gap is observed. As the
structural/magnetic transitions are separated in the
Na$_{1-\delta}$FeAs case, we find the SDW gap opens in accordance
with the magnetic transition. Comparing with the optical response
of other FeAs-based parent compounds, both the gap value 2$\Delta$
and the energy scale for the gap-induced spectral weight
redistribution are smaller in Na$_{1-\delta}$FeAs. Our findings
support the itinerant origin of the antiferromagnetic transition
in the FeAs-based system.

\end{abstract}



\maketitle

The interplay between different instabilities, such as structural
distortions, magnetic orderings, and superconductivity, is of
central interest in condensed matter physics. The discovery of
superconductivity in FeAs based layered materials
\cite{Kamihara08} offers a new opportunity to study the intriguing
interplay between those instabilities. The undoped FeAs-based
compounds commonly display the structural and magnetic phase
transitions, which, depending on materials, could occur either at
the same temperature or separately.\cite{Cruz,Rotter2,Chu} The
magnetic order has a collinear spin structure with a ($\pi$,
$\pi$) wavevector in the folded Brillioun zone (two Fe ions per
unit cell). Upon electron or hole doping or application of
pressure, both the magnetic order and the structural transition
are suppressed, and superconductivity
emerges.\cite{JZhao,Rotter2,Torikachvili} It is widely believed
that the structural distortion is driven by the magnetic
transition,\cite{Yildirim,FangXu} however, there have been much
debate on whether the parent compounds belong to the local or
itinerant category of antiferromagnets. One pool of theories
invoke an itinerant electron approach to the systems in which the
commensurate antiferromagnetic (AFM) order originates from a
spin-density wave (SDW) instability due to the the nesting of the
electron and hole Fermi surfaces which are separated by a ($\pi$,
$\pi$) wavevector.\cite{Dong,Mazin,Ran,Tesanovic,SinghReview} The
itinerant electrons and holes are removed by the gapping of the
Fermi surface (FS) in the SDW ordered state. Alternatively, a
Heisenberg magnetic exchange model is suggested to explain the AFM
structure.\cite{Yildirim,Si,Ma,FangXu,Wu} In this picture, the AFM
order is a signature of local physics. On the other hand, Igor
Mazin recently argued that neither the itinerant nor the local
moment pictures are fully correct. The moments are largely local,
driven by the Hund's coupling rather than by the on-site Hubbard
repulsion, while the ordering is driven mostly by the gain of the
one-electron energies of all occupied states.\cite{Mazin3}

The advantage for spectroscopic techniques is detecting the energy
gap in the broken symmetry state. Previous optical investigations
provide clear evidence for the formation of the SDW partial gap in
the magnetic phase in polycrystalline ReFeAsO (Re=La, Ce, Nd,
etc)\cite{Dong,ChenPRLCe,HuReview,Boris} and single crystalline
AFe$_2$As$_2$ (A=Ba, Sr)\cite{Hu122,DWu} which, therefore, support
the itinerant picture that the energy gain for the AFM ground
state is achieved by the opening of an SDW gap on the Fermi
surface. Quantum oscillation experiment revealed three Fermi
surfaces with area much smaller than those in the paramagnetic
phase predicted by the local density approximation
calculation,\cite{Sebastian} thus agree with optical observation
of a large reduction of effective carrier density in the SDW
state. However, angle-resolved photoemission spectroscopy (ARPES)
experiments did not yield consistent
results.\cite{Kaminski,Feng,Hsieh,GDLiu} On the other hand, the
iron chalcogen-based parent compound, Fe$_{1+x}$Te, which also
exhibits structural and magnetic phase transition near 65 K, shows
no signature of gap opening for the magnetic ordered state from
infrared spectroscopy measurement.\cite{Chen2} As neutron
experiments revealed that the low-\emph{T} magnetic phase has a
bi-collinear spin structure with a ($\pi$, 0)
wavevector,\cite{Bao,Li} which is different from the ($\pi$,
$\pi$) wavevector that connects the electron and hole pockets, the
absence of gap opening below \emph{T$_{SDW}$} is not surprising.
Anyhow, further spectroscopic studies are required to find out
whether the gap formation is a common feature for different types
of Fe-based parent compounds. Another important question is
whether the gap emerges after the structural distortion or below
the magnetic transition. As the magnetic and structural
transitions occur simultaneously in AFe$_2$As$_2$ (122-type),
while no single crystal with sufficient size for optical
measurement is obtained for ReFeAsO (1111-type), finding a new
type of FeAs-based parent compound with well separated
structural/magnetic transitions from which sizeable single
crystals can be easily obtained is highly required.

Recently, high quality single crystals are synthesized for almost
stoichiometric Na$_{1-\delta}$FeAs, a 111-type FeAs-based parent
compound. Two separated structural/magnetic transitions at 52 and
41 K, together with a superconducting transition at 23 K are found
by transport measurements.\cite{ChenNaFeAs} Here we report the
in-plane optical properties for Na$_{1-\delta}$FeAs single
crystal. Clear optical evidence for the SDW gap is found, and the
gap emerges in accordance with the magnetic transition. Moveover,
the gap value 2$\Delta$ has a smaller energy scale than 122-type
compounds with higher SDW transition temperatures. The ratio of
2$\Delta$/$k_BT_{SDW}\approx$4.2 is close to the expectation of
mean field theory for an itinerant SDW order. Like the 122
systems, a residual Drude term (free-carrier response) is seen in
the SDW ordered phase, thus Na$_{1-\delta}$FeAs is still metallic
when the SDW gap develops. Then, metallic response and the SDW gap
appear to be a common feature for the undoped FeAs-based
materials.

\begin{figure}[t]
\includegraphics[width=2.7in]{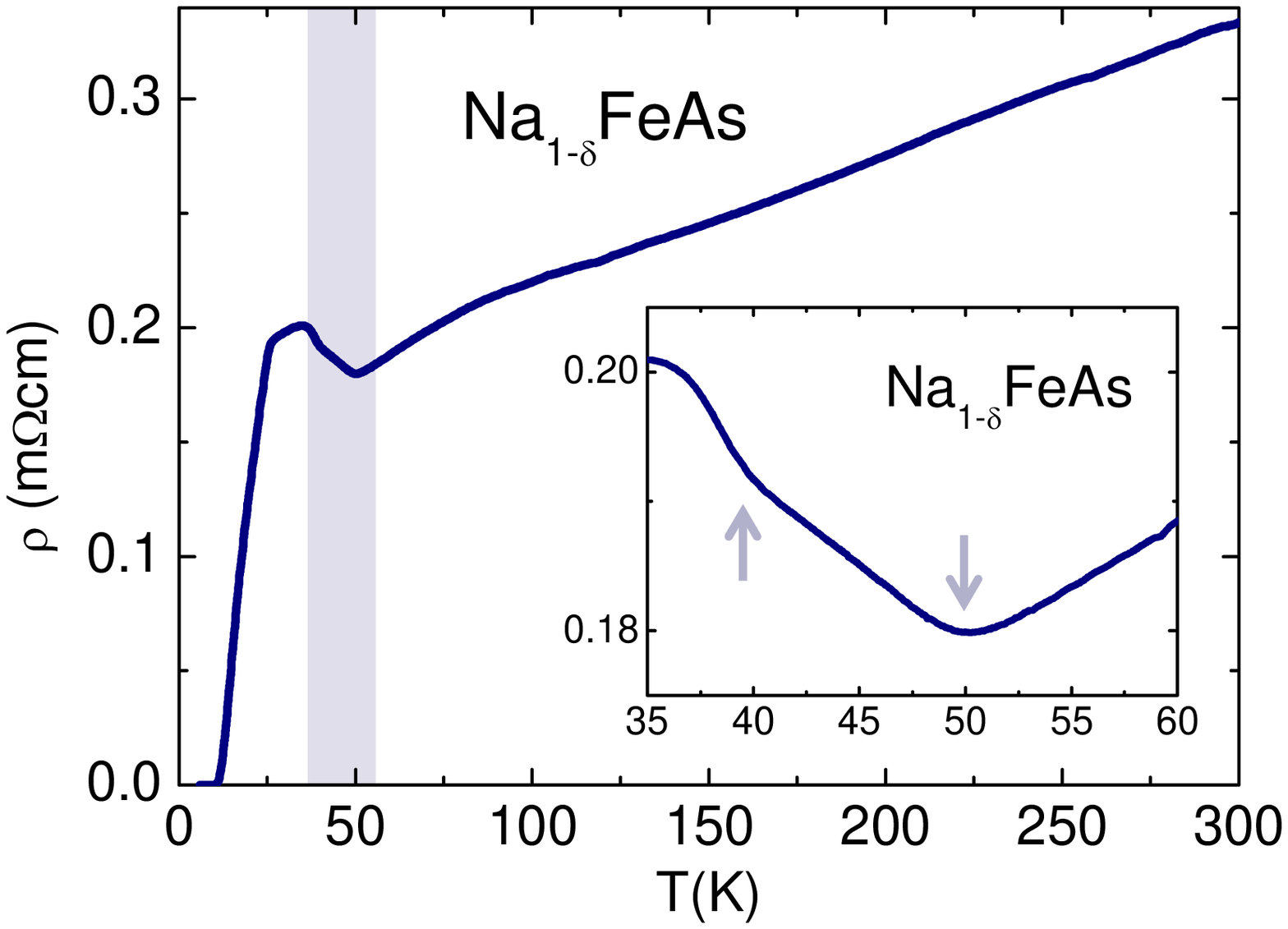}
\caption{The in-plane \emph{dc} resistivity for
Na$_{1-\delta}$FeAs single crystal. Two separated transitions
around 40 and 50 K can be better resolved in the inset
figure.\label{fig:rho}}
\end{figure}

Single crystalline Na$_{1-\delta}$FeAs samples were grown by the
self-flux method.\cite{ChenNaFeAs} The obtained crystals can be
easily cleaved along ab-plane. The \emph{dc} resistivity $\rho(T)$
is obtained by the standard four-probe method on a sample cleaved
from the same crystal used in the optical measurement. The result
is shown in Fig.\ref{fig:rho}. Two transitions near 40 and 50 K
were assigned to separated structural and magnetic
transitions,\cite{ChenNaFeAs} which were confirmed by recent
neutron diffraction measurement.\cite{LiNaFeAs} Here the \emph{dc}
resistivity turns up after the structural distortion, and
increases more rapidly with decreasing \emph{T} in the SDW state.
A superconducting transition is seen with an onset temperature of
25 K, and a zero resistivity is approached when T$\leq$10 K. This
was interpreted as due to the slight Na deficiency (less than
1\%). Since no detectable specific heat anomaly around
\emph{T$_c$} was found, the superconducting volume fraction is
rather small.\cite{ChenNaFeAs} We expect that the optical data are
dominated by the response of the parent phase.

The optical reflectance measurements were performed on a
combination of Bruker IFS 66v/s and 113v spectrometers on newly
cleaved surfaces (ab-plane) in the frequency range from 40 to
25000 cm$^{-1}$. An \textit{in situ} gold and aluminium
overcoating technique was used to get the reflectivity
R($\omega$). The real part of conductivity $\sigma_1(\omega)$ is
obtained by the Kramers-Kronig transformation of R($\omega$).

\begin{figure}[t]
\includegraphics[width=3.2in]{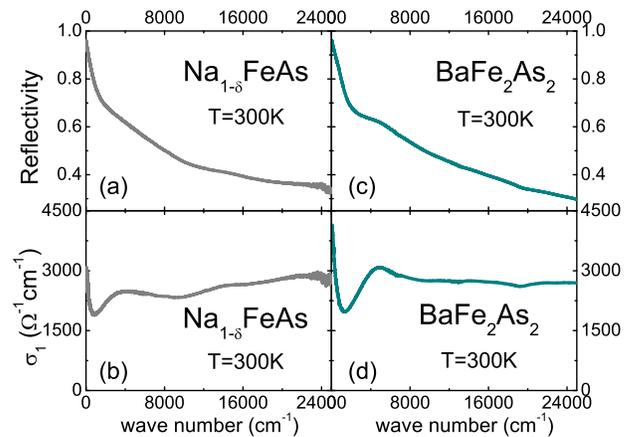}
\caption{The room-temperature optical reflectivity (a) and
conductivity (b) for Na$_{1-\delta}$FeAs single crystal over broad
frequencies up to 25000 \cm. The optical data on BaFe$_2$As$_2$
crystal \cite{Hu122} in the same frequency range are shown in (c)
and (d) for comparison.\label{fig:RScmp}}
\end{figure}

Figure \ref{fig:RScmp} shows the room-temperature optical
reflectivity and conductivity spectra over broad frequencies up to
25000 \cm. The overall spectral lineshapes are very similar to the
AFe$_2$As$_2$ (A=Ba, Sr) single crystals. As a comparison, we have
included the optical data on BaFe$_2$As$_2$ in the
figure.\cite{Hu122} The reflectance drops almost linearly with
frequency at low-$\omega$ region, then merges into the high values
of a background contributed mostly from the interband transitions
from the mid-infrared to visible regime. By fitting the
conductivity spectrum with the Drude and Lorentz model in a way
similar to what we did for 122-type crystals,\cite{Hu122} we get
the plasma frequency $\omega_p$$\approx$10200 \cm and scattering
rate $1/\tau$$\approx$650 \cm for Na$_{1-\delta}$FeAs. Both are
comparable to the parameters found for the 122-type
materials.\cite{Hu122,DWu} This indicates that we are measuring
the charge dynamics of Fe$_2$As$_2$ layers.

\begin{figure}[b]
\includegraphics[width=3.2in]{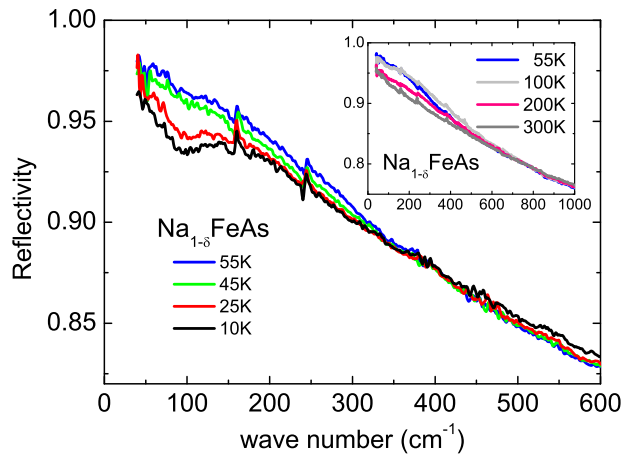}
\caption{(Color online) Optical reflectivity R($\omega$) for
Na$_{1-\delta}$FeAs single crystal. The main figure shows
R($\omega$) below 600 \cm for \emph{T}$\leq$ 55 K. The SDW gap is
evidenced by a spectral suppression in the far-infrared for
\emph{T}=10 and 25 K. The inset plots R($\omega$) below 1000 \cm
for 55, 100, 200 and 300 K.\label{fig:R}}
\end{figure}

Figure \ref{fig:R} shows the optical reflectivity R($\omega)$ for
Na$_{1-\delta}$FeAs below 600 \cm. Two phonons around 160 and 244
\cm can be found. R($\omega)$ at low \emph{T} decreases gradually
below 400 \cm, consists with the increasing of $\rho(T)$ below 50
K. In the SDW state, a suppression in R($\omega$) below 150 \cm is
observed, indicating the opening of an SDW gap on the Fermi
surface. Similar cases were found for the undoped
122-type\cite{Hu122,DWu,Pfuner} and 1111-type\cite{HuReview,
Boris} compounds. Here the suppression exists only for \emph{T}=10
and 25 K, while no clear indication for the gap is seen at
\emph{T}=45 K when the sample just experiences a structural
distortion but without any magnetic ordering. The inset plots the
low frequency R($\omega$) for \emph{T}=55, 100, 200 and 300 K.
Here R($\omega)$ shows a metallic response in the far-infrared
region, that the reflectivity continues to grow with lowering
\emph{T} in the normal state, in agreement with the metallic
response as seen in the \emph{dc} resistivity.

\begin{figure}[t]
\includegraphics[width=3.4in]{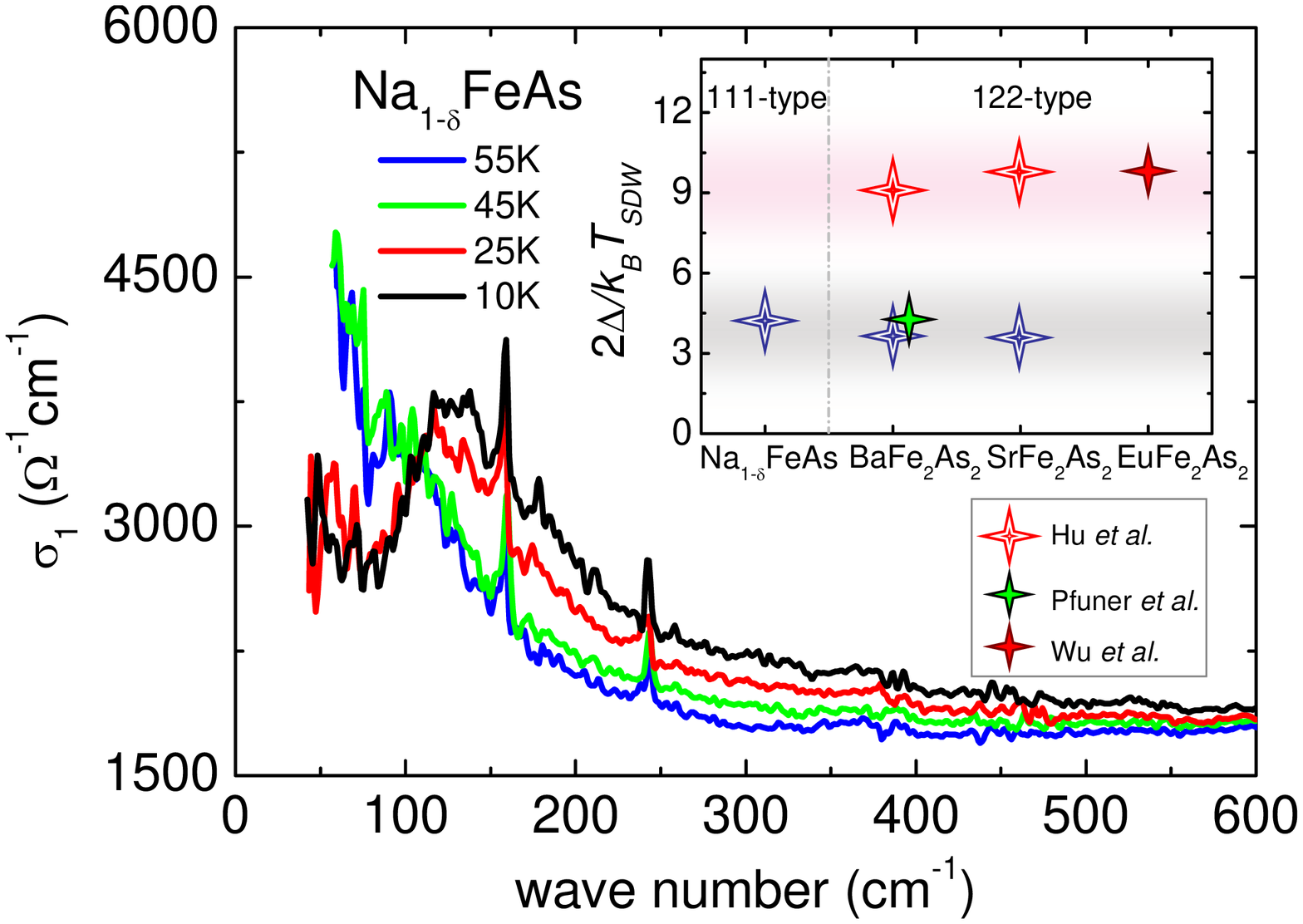}
\caption{(Color online) The real part of optical conductivity
$\sigma_1(\omega)$ for Na$_{1-\delta}$FeAs below 600 \cm. A SDW
gap feature emerges for \emph{T}$<$\emph{T$_{SDW}$}. Inset:
2$\Delta$/$k_B$$T_{SDW}$ obtained from optical data for various
single crystalline FeAs-based parent compounds, including
Na$_{1-\delta}$FeAs (this study), BaFe$_2$As$_2$ (Hu \emph{et
al.}\cite{Hu122}, Pfuner \emph{et al.}\cite{Pfuner}),
SrFe$_2$As$_2$ (Hu \emph{et al.}\cite{Hu122}), and EuFe$_2$As$_2$
(Wu \emph{et al.}\cite{DWu}).\label{fig:S1}}
\end{figure}

The low-$\omega$ temperature-dependent real part of conductivity
$\sigma_1(\omega)$ is shown in Fig. \ref{fig:S1}. Besides two
sharp phonon modes around 160 and 244 \cm, $\sigma_1(\omega)$ for
both \emph{T}=45 and 55 K show a Drude response without any clear
evidence for the gap-induced absorption peak. For \emph{T}=10 and
25 K, a peak with a clear edge-like feature is formed around 120
\cm. Therefore, the energy gap emerges only in the
antiferromagnetic state. Meanwhile, a remaining Drude component is
seen below 80 \cm, indicating that the Fermi surface is only
partially gapped and Na$_{1-\delta}$FeAs is still metallic in the
SDW state. In our earlier study on AFe$_2$As$_2$ (A=Ba, Sr) single
crystals, we know that the residual Drude component in the SDW
state has a much smaller spectral weight and a narrower peak
width, indicating the removal of both conducting carriers and the
scattering channel.\cite{Hu122} Here, essentially we see the same
structural feature. However, because of the limited frequency
range below the energy gap, a quantitative estimation for the loss
of carrier density and scattering rate could not be accurately
determined.

The optical conductivity $\sigma_1(\omega)$ shows different gap
characters for the superconducting and the density wave states due
to their respective coherence factors.\cite{Gruner,Tinkham} For
the SDW ground state with an isotropic gap, a non-symmetric peak
with clear edge-like feature emerges at 2$\Delta$ in the optical
conductivity, so that $\sigma_1^{SDW}(\omega)$ exceeds the normal
state conductivity $\sigma_1^{N}(\omega)$ at the gap
onset,\cite{Gruner} above which the loss in free-carrier (Drude)
spectral weight is gradually compensated by the gap-induced
absorption peak. For a multi-band system, gap anisotropy will
weaken the edge-like feature at 2$\Delta$. Here we use the peak
position (the conductivity maximum) to estimate the SDW gap. For
Na$_{1-\delta}$FeAs, the conductivity peak at 10 K is around 120
\cm, i.e., 2$\Delta_{SDW}$$\approx$15 meV. Such a gap is obviously
smaller than that in AFe$_2$As$_2$ and
ReFeAsO.\cite{Hu122,HuReview,DWu,Boris,Pfuner}

In undoped 122-type compounds AFe$_2$As$_2$, (A=Sr, Ba, Eu), a
double-gap character is found with the
2$\Delta$/$k_B$$T_{SDW}\sim$ 9 and 4, respectively. For
Na$_{1-\delta}$FeAs, only one energy gap feature is observed in
the measurement frequency range with
2$\Delta$/$k_B$$T_{SDW}$$\sim$ 4.2. From the spectral lineshape,
this gap feature should correspond to the higher energy gap
feature in undoped 122 compounds. It is not clear whether a second
feature at lower energy scale exists beyond the lowest measurement
energy. Thus, although qualitatively the energy gaps are smaller
for undoped compounds with lower SDW transition temperatures,
there is no scaling relation between different compounds with
different $T_{SDW}$. In the inset of Fig. \ref{fig:S1} we show
2$\Delta$/$k_B$$T_{SDW}$ obtained by optical data on
111(Na$_{1-\delta}$FeAs) and 122 (AFe$_2$As$_2$,
A=Sr,Ba,Eu)\cite{Hu122,Pfuner,DWu} type parent compounds. Here the
gap 2$\Delta$ for different FeAs systems are all defined by the
peak positions in $\sigma_1(\omega)$ for consistency.

\begin{figure}[b]
\includegraphics[width=3.2in]{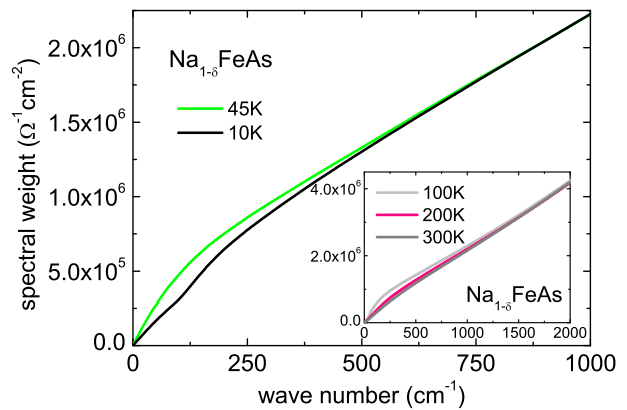}
\caption{(Color online) The spectral weight $\int_{\rm{0}}^\omega
{\sigma _{\rm{1}} (\omega )} d\omega$ for Na$_{1-\delta}$FeAs
below 1000 \cm for \emph{T}=10 and 45 K. Inset: the spectral
weight for \emph{T}=100, 200, and 300 K up to 2000
\cm.\label{fig:SW}}
\end{figure}

Besides the gap value, the energy scale affected by the SDW gap is
also smaller for Na$_{1-\delta}$FeAs in comparison with other
FeAs-based parent compounds with higher \emph{T$_{SDW}$}s. Figure
\ref{fig:SW} plots the spectral weight at 10 and 45 K for
Na$_{1-\delta}$FeAs. The inset shows the spectral weight at
\emph{T}=100, 200 and 300 K. Note Na$_{1-\delta}$FeAs is metallic
in the normal state (Fig.\ref{fig:rho}), so the spectral weight
piles up at low frequencies with decreasing \emph{T}, which is due
to an increasing \emph{dc} conductivity ($\sigma_1(0)$) thus a
growing Drude peak in $\sigma_1(\omega)$. Above 1700 \cm, the
spectral weight for all temperatures merge together. In the SDW
state, the spectral weight is smaller for 10 K than that of 45 K,
indicating a loss in the Drude weight at 10 K, that part of the
free carriers are removed from E$_F$ due to the SDW gap. The
spectral weight loss in low frequencies is compensated when
$\omega$ approaches 750 \cm. Such an energy scale is smaller than
that of AFe$_2$As$_2$ (e.g. 2000 \cm for BaFe$_2$As$_2$ where
T$_{SDW}$$\approx$140 K).\cite{Hu122}

Our study clearly indicates that the metallic response and the
opening of an energy gap in the magnetic ordered state are
ubiquitous behaviors for all FeAs-based undoped compounds. In
addition, the gap magnitude correlates with \emph{T$_{SDW}$}.
Associated with the gapping of the Fermi surface, a large part of
the Drude component is removed (indicating a reduction of the FS
area) and the scattering channel is also reduced. All favor an
itinerant origin of the SDW order. We noticed that some ARPES
studies\cite{Feng,GDLiu} on BaFe$_2$As$_2$ did not reveal any gap
in the SDW state, therefore failed to see a dramatic reduction of
Fermi surface areas. As the optics probes the bulk properties,
while the ARPES is mainly a surface probe, it remains to clarify
if there is a surface reconstruction which would affect the
results. Considering the multi-band/orbital character for
FeAs-based compounds, the entire band structure might be
reconstructed when parts of them were modified by FS nesting
instability. In this sense, our optical data do not conflict with
the band reconstruction picture.

Finally, we comment on the upturn behavior of the \emph{dc}
resistivity below the structural/magnetic phase transition. Note
$\rho(T)$ turns up for Na$_{1-\delta}$FeAs but drops more steeply
with decreasing \emph{T} for AFe$_2$As$_2$ and ReFeAsO. From the
semi-classic Boltzmann transport theory, the resistivity is
determined by the complex function of Fermi velocity, the
scattering rate, and their weighted integral over the whole
FS.\cite{Dressel} In the case of electron gas, it could be
simplified to the Drude form for which the resistivity is
determined by carrier density and scattering rate. So apparently,
whether $\rho$ shows an upturn or a fast drop depends on the
subtle balance of those quantities which experience substantial
changes across the transition.

To summarize, we studied the in-plane optical properties for
Na$_{1-\delta}$FeAs, a FeAs-based parent compound with separated
structural and magnetic transitions. It shares similar optical
response over broad frequencies with other FeAs-based systems. A
clear energy gap in $\sigma_1(\omega)$ is observed below the
magnetic phase transition, accompanied by a spectral weight
transfer from the free-carrier Drude term to above this gap
energy. Both the gap 2$\Delta$ and the energy scale associated
with the spectral weight redistribution are smaller in comparison
with other undoped FeAs-based compounds with higher
\emph{T$_{SDW}$}. The results favor an itinerant origin for the
SDW transition.

This work is supported by the NSFC, CAS, and the 973 project of
the MOST of China.


\begin{thebibliography}{99}

\bibitem{Kamihara08} Y. Kamihara, T. Watanabe, M. Hirano, and
H. Hosono, J. Am. Chem. Soc. 130, 3296 (2008).

\bibitem{Cruz}  Clarina de la Cruz, Q. Huang, J. W. Lynn, Jiying Li, W. Ratcliff II, J. L. Zarestky, H. A. Mook, G. F. Chen,
J. L. Luo, N. L. Wang, and Pengcheng Dai, Nature \textbf{453}, 899
(2008).

\bibitem{Rotter2} M. Rotter, M. Tegel, D. Johrendt, Phys. Rev. Lett. \textbf{101}, 107006 (2008).

\bibitem{Chu} J.H. Chu, J.G. Analytis, C. Kucharczyk, and I.R.
Fisher, Phys. Rev. B \textbf{79} 014506 (2009).

\bibitem{JZhao}Jun Zhao, Q. Huang, Clarina de la Cruz, Shiliang Li, J. W. Lynn,
Y. Chen, M. A. Green, G. F. Chen, G. Li, Z. Li, J. L. Luo, N. L.
Wang, Pengcheng Dai,  Nature Materials \textbf{7}, 953 (2008).

\bibitem{Torikachvili}  M. S. Torikachvili, S. L. Bud'ko, N. Ni, P. C. Canfield, Phys. Rev. Lett. \textbf{101},
057006 (2008).

\bibitem{Yildirim} T. Yildirim, Phys. Rev. Lett. \textbf{101}, 057010 (2008).

\bibitem{FangXu} C. Fang, H. Yao, W.-F. Tsai, J. P. Hu, S. A. Kivelson,
Phys. Rev. B \textbf{77}, 224509 (2008); C. Xu, M. Mueller, S.
Sachdev, Phys. Rev. B \textbf{78}, 020501(R) (2008).

\bibitem{Dong} J. Dong, H. J. Zhang, G. Xu, Z. Li, G. Li, W. Z. Hu, D. Wu,
G. F. Chen, X. Dai, J. L. Luo, Z. Fang, N. L. Wang, Europhys.
Lett. \textbf{83}, 27006 (2008).

\bibitem{Mazin} I.I. Mazin, D.J. Singh, M.D. Johannes, M.H. Du, Phys. Rev. Lett. \textbf{101}, 057003
(2008).

\bibitem{Ran} Ying Ran, Fa Wang, Hui Zhai, Ashvin Vishwanath, Dung-Hai Lee, Phys. Rev. B \textbf{79}, 014505 (2009).

\bibitem{Tesanovic} V. Cvetkovic and Z. Tesanovic, Europhysics Letters \textbf{85}, 37002 (2009).

\bibitem{SinghReview} D. J. Singh, Physica C \textbf{469}, 418 (2009) and references therein.

\bibitem{Si} Q. Si and E. Abrahams, Phys. Rev. Lett. \textbf{101}, 076401 (2008).

\bibitem{Ma} F. J. Ma, Z.Y. Lu, and T. Xiang, Phys. Rev. B \textbf{78}, 224517 (2008).

\bibitem{Wu} J. Wu, P. Phillips and A. H. Castro Neto,
Phys. Rev. Lett. \textbf{101}, 126401 (2008).


\bibitem{Mazin3} M. D. Johannes and I. I. Mazin, Phys. Rev. B \textbf{79}, 220510(R)
(2009).

\bibitem{ChenPRLCe} G. F. Chen, Z. Li, D. Wu, G. Li, W. Z. Hu, J. Dong, P. Zheng, J. L. Luo, and N. L.
Wang, Phys. Rev. Lett. \textbf{100}, 247002 (2008).

\bibitem{HuReview} W. Z. Hu, Q. M. Zhang, and N. L. Wang, Physica C
\textbf{469}, 545 (2009).

\bibitem{Boris} A.V. Boris, N. N. Kovaleva, S. S. A. Seo, J. S. Kim, P.
Popovich, Y. Matiks, R. K. Kremer, and B. Keimer, Phys. Rev. Lett.
\textbf{102}, 027001 (2009).


\bibitem{Hu122} W. Z. Hu, J. Dong, G. Li, Z. Li, P. Zheng, G. F. Chen, J. L. Luo, and N. L.
Wang, Phys. Rev. Lett. \textbf{101}, 257005(2008).

\bibitem{DWu} D. Wu, N. Bari\v{s}i\'{c}, N. Drichko, S. Kaiser, A. Faridian, M.
Dressel, S. Jiang, Z. Ren, L. J. Li, G. H. Cao, Z. A. Xu, H. S.
Jeevan and P. Gegenwart, Phys. Rev. B \textbf{79}, 155103 (2009).

\bibitem{Sebastian} Suchitra E. Sebastian, J. Gillett, N. Harrison, P. H. C. Lau,
D. J. Singh, C. H. Mielke and G. G. Lonzarich, J. Phys.: Condens.
Matter \textbf{20}, 422203 (2008).

\bibitem{Kaminski} C. Liu, G. D. Samolyuk, Y. Lee, N. Ni, T. Kondo, A. F.
Santander-Syro, S. L. Bud'ko, J. L. McChesney, E. Rotenberg, T.
Valla, A. V. Fedorov, P. C. Canfield, B. N. Harmon, and A.
Kaminski, Phys. Rev. Lett. \textbf{101}, 177005 (2008).

\bibitem{Feng} L. X. Yang, Y. Zhang, H. W. Ou, J. F. Zhao, D. W. Shen, B. Zhou,
J. Wei, F. Chen, M. Xu, C. He, Y. Chen, Z. D. Wang, X. F. Wang, T.
Wu, G. Wu, X. H. Chen, M. Arita, K. Shimada, M. Taniguchi, Z. Y.
Lu, T. Xiang, and D. L. Feng, Phys. Rev. Lett. \textbf{102},
107002 (2009).

\bibitem{Hsieh} D. Hsieh, Y. Xia, L. Wray, D. Qian, K. K. Gomes,
A. Yazdani, G. F. Chen, J. L. Luo, N. L. Wang, and M. Z. Hasan,
arXiv:0812.2289.

\bibitem{GDLiu} Guodong Liu, Haiyun Liu, Lin Zhao, Wentao Zhang, Xiaowen Jia, Jianqiao Meng, Xiaoli Dong, G. F. Chen,
Guiling Wang, Yong Zhou, Yong Zhu, Xiaoyang Wang, Zuyan Xu,
Chuangtian Chen, and X. J. Zhou, arXiv:0904.0677v2.

\bibitem{Chen2} G. F. Chen, Z. G. Chen, J. Dong, W. Z. Hu, G. Li, X. D. Zhang, P. Zheng, J. L. Luo,
and N. L. Wang,  Phys. Rev. B \textbf{79}, 140509(R) (2009).

\bibitem{Bao} W. Bao, Y. Qiu, Q. Huang, A. Green, P. Zajdel,
M.R. Fitzsimmons, M. Zhernenkov, M. Fang, B. Qian, E.K. Vehstedt,
J. Yang, H.M. Pham, L. Spinu, and Z.Q. Mao, arXiv:0809.2058.

\bibitem{Li} Shiliang Li, Clarina de la Cruz, Q. Huang, Y. Chen, J. W. Lynn,
Jiangping Hu, Yi-Lin Huang, Fong-chi Hsu, Kuo-Wei Yeh, Maw-Kuen
Wu, and Pengcheng Dai, Phys. Rev. B \textbf{79}, 054503 (2009).

\bibitem{ChenNaFeAs} G. F. Chen, W. Z. Hu, J. L. Luo, and N. L.
Wang, Phys. Rev. Lett. \textbf{102}, 227004 (2009).

\bibitem{LiNaFeAs} Shiliang Li, Clarina de la Cruz, Q. Huang, G. F. Chen, T.-L. Xia, J. L. Lou, N.
L. Wang, Pengcheng Dai, arXiv:0905.0525 (Phys. Rev. B, Rapid
Commun.(in-press, 2009)).

\bibitem{Pfuner} F. Pfuner, J.G. Analytis, J.-H. Chu, I.R. Fisher, and L. Degiorgi,
Eur. Phys. J B \textbf{67}, 513 (2009).

\bibitem{Gruner} G. Gr\"{u}ner, \emph{Density Waves in Solids} (Addison-Weslsy, Reading, MA,
1994).

\bibitem{Tinkham} M. Tinkham, \emph{Introduction to Superconductivity}(2nd Ed.), (McGraw-Hill, New
York, 1996).

\bibitem{Dressel} M. Dressel, and G. Gr\"{u}ner, \emph{Electrodynamics of Solids}, Cambridge University Press (2002).

\end{thebibliography}
\end{document}